\documentclass[superscriptaddress,secnumarabic,amssymb,amsmath,nobibnotes,aps,prd,showkeys,showpacs,nofootinbib]{revtex4}%
\usepackage{graphicx}
\usepackage{epsf}
\usepackage{bm}
\usepackage{amsmath}
\usepackage{amsfonts}
\usepackage{amssymb}%
\providecommand{\U}[1]{\protect\rule{.1in}{.1in}}
\newcommand{\be}{\begin{equation}}
\newcommand{\ee}{\end{equation}}

\newcommand{\mincir}{\raise
-3.truept\hbox{\rlap{\hbox{$\sim$}}\raise4.truept\hbox{$<$}\ }}
\newcommand{\magcir}{\raise
-3.truept\hbox{\rlap{\hbox{$\sim$}}\raise4.truept\hbox{$>$}\ }}

\begin{document}
\title{Conformally related metrics and Lagrangians  and their physical interpretation in cosmology}
\author{Michael Tsamparlis}
\affiliation{Faculty of Physics, Department of Astrophysics - Astronomy - Mechanics
University of Athens, Panepistemiopolis, Athens 157 83, Greece}
\author{Andronikos Paliathanasis}
\affiliation{Faculty of Physics, Department of Astrophysics - Astronomy - Mechanics
University of Athens, Panepistemiopolis, Athens 157 83, Greece}
\author{Spyros Basilakos}
\affiliation{Academy of Athens, Research Center for Astronomy and Applied Mathematics,
Soranou Efesiou 4, 11527, Athens, Greece}
\author{Salvatore Capozziello}
\affiliation{Dipartimento di Fisica, Universita' di Napoli "Federico II"}
\affiliation{and INFN Sez. di Napoli, Compl. Univ. di Monte S. Angelo, Ed. G., Via Cinthia,
9, I-80126, Napoli, Italy.}

\begin{abstract}
Conformally related metrics and Lagrangians are considered
in the context of scalar-tensor gravity  cosmology. After the discussion of the problem,  we pose a lemma in which
we show that the field equations of two conformally related Lagrangians are
also conformally related if and only if the corresponding Hamiltonian
vanishes. Then we prove that to every non-minimally coupled scalar field, we
may associate a unique minimally coupled scalar field in a conformally related
space with an appropriate potential. The latter result implies that the field
equations of a non-minimally coupled scalar field are the same at the
conformal level with the field equations of the minimally coupled scalar
field. This fact  is relevant in order to select physical variables among conformally equivalent systems. Finally, we find that the above propositions can be extended to a
general Riemannian space of $n$-dimensions. 

\end{abstract}
\date{\today}

\maketitle

\hyphenation{tho-rou-ghly in-te-gra-ting e-vol-ving con-si-de-ring
ta-king me-tho-do-lo-gy fi-gu-re}

\section{Introduction}
The detailed analysis of the current high
quality cosmological data (Type Ia supernovae, cosmic microwave background,
baryonic acoustic oscillations, etc), converge towards  a new  emerging ``Cosmological Standard
Model''. This cosmological model is spatially flat with a cosmic dark sector
constituted  by cold dark matter and some sort of dark energy, associated
with large negative pressure, in order to explain the observed accelerating
expansion of the universe (see
\cite{Teg04,Spergel07,essence,Kowal08,Hic09,komatsu08,LJC09,BasPli10} and
references therein). Despite the mounting observational evidences on the
existence of the dark energy component in the universe, its nature and
fundamental origin remains an intriguing enigma challenging the very
foundations of theoretical physics.

Indeed, during the last decade there has been an intense theoretical debate
among cosmologists regarding the nature of this exotic ``dark energy''. The
absence of a fundamental physical theory, concerning the mechanism inducing
the cosmic acceleration, has opened a window to a plethora of alternative
cosmological scenarios. Most are based either on the existence of new fields
in nature (dark energy) or in some modification of Einstein's general
relativity, with the present accelerating stage appearing as a sort of
geometric effect (see \cite{Ratra88,curvature,mauro,report,repsergei,
Oze87,Weinberg89,Lambdat,Bas09c,Wetterich:1994bg,
Caldwell98,Brax:1999gp,KAM,fein02,Caldwell,Bento03,chime04,Linder2004,LSS08,
Brookfield:2005td,Boehmer:2007qa,Starobinsky-2007,Ame10} and references therein).

In order to investigate the dynamical properties of a particular
\textquotedblleft dark energy\textquotedblright\ model, we need to specify the
covariant Einstein-Hilbert action of the model and find out the corresponding
energy-momentum tensor. This methodology provides a mathematically consistent
way to incorporate \textquotedblleft dark energy\textquotedblright\ in
cosmology. However, in the literature, there are many Lagrangians
\cite{Ame10} which describe differently the physical features of the scalar
field or the modified gravity \cite{report}. Because of the large amount of
\textquotedblleft dark energy\textquotedblright\ models, it is essential
to study them in a unified context in order to discriminate the true physical variables. From our viewpoint,  this framework
has to be at the level of geometry since the various Lagrangians which
describe the nature of \textquotedblleft dark energy\textquotedblright\ are
embedded in the space-time. From a theoretical point of view, an easy way to
study  dynamics in a unified manner is to look for conformally related
Lagrangians. In fact, the idea to use conformally related metrics and
Lagrangians as a cosmological tool is not new. In particular, it has been
proposed that the existence of conformally equivalent Lagrangians can be used
in order to select viable cosmological models \cite{allemandi}.

In general, the presence of scalar fields into the gravitational action can 
give rise to two classes of theories: minimally and non-minimally coupled Lagrangians.
In the first case, the gravitational coupling is the standard Newton constant and the scalar field 
Lagrangian is simply added to the Ricci scalar. It can consist of $i)$ a kinetic term,  
$ii)$ a kinetic term and a self-interaction potential, or $iii)$ just an interaction potential. In the first case, the 
scalar field is nothing else but a cyclic variable and then it is related to a conserved quantity. 
The second case is the relevant one since, by a variational principle, it is possible to obtain a Klein-Gordon equation where 
the self-interacting potential  $V(\phi)$ leads the dynamics. The third case means that the scalar field has no dynamics. 
When the coupling is the Newton constant, we are in the so-called {\it Einstein frame}.

In fact, as firstly pointed out by Brans and Dicke \cite{brans}, a gravitational theory can be made more "Machian"  
by relaxing the hypothesis that the coupling is constant. 
They introduced a scalar field $\phi$ non-minimally coupled to the Ricci scalar $R$ and a kinetic term for 
such a scalar field into the gravitational action. The result was that the coupling was non-minimal and coordinate dependent. 
In other words, the gravitational interaction is assumed to change with distance and time. This approach can 
be generalized considering scalar-tensor theories of gravity where also a  self-interaction potential comes into 
the game or more than one scalar field are  taken into account. In general, any gravitational theory not simply 
linear in the Ricci scalar  can be reduced to a scalar-tensor one.  In the case of $f(R)$-gravity, it is straightforward to 
show that an  {\it O'Hanlon representation} by scalar fields is possible \cite{ohanlon}.  In this case, a scalar field is non-minimally coupled to the Ricci 
scalar and a self-interacting potential is present  while there is no kinetic term. Scalar field dynamics is guaranteed by 
the non-minimal coupling and the potential (see \cite{report} and references therein). In other words, the further 
gravitational degrees of freedom, coming from the fact that $f(R)\neq R$, can be figured out as a scalar field.    
As soon as we are considering non-minimal couplings or higher-order terms in the Lagrangian, we are in 
the so-called {\it Jordan frame}, after Jordan who first introduced this notion \cite{Jordan38,Jordan52}. 

In this article we would like to address the following basic question: \textit{In the framework of the scalar field
cosmology, is it possible to relate the available Lagrangians in a conformal
way?} 

The structure of the paper is as follows. 
In section II, we discuss the issue of conformal transformations considering its historical development and connection with physical theories. 
From our point of view, this section is essential in order to fix the problem showing the urgency to discriminate physical variables among the conformally related models.
The basic theoretical elements
of the conformally related metric are presented in section III, where we also
introduce the concept of conformally related Lagrangians and prove a lemma
which shows that the field equations for two conformally related Lagrangian
are also conformally related if the corresponding Hamiltonian vanishes. In
section IV, we discuss the conformal equivalence of Lagrangians for scalar
fields in a Riemannian space of dimension $4$ (for an extension see appendix
A). In particular we enunciate  a theorem which proves that the field equations for
a non-minimally coupled scalar field are the same at the conformal level with
the field equations of the minimally coupled scalar field. The necessity to
preserve Einstein's equations in the context of Friedmann-Lema\^\i
tre-Robertson-Walker (FRLW) space-time leads us to apply, in section V, the
current general analysis to the scalar field (quintessence or phantom) flat
FLRW cosmologies. Finally, we draw our main conclusions  and discuss results in section VI.

\section{What is the physical frame?}

Some considerations are in order at this point. The conformal
transformations from the Jordan to the Einstein frame are geometrical maps
allowing to set many features of scalar-tensor gravity, $f(R)$ gravity and, in
general, any modified theory of gravity. Taking into account both the Jordan
and the Einstein conformal frames (infinite conformal frames can be chosen
assuming a suitable conformal factor), the question is whether the frames are infinitely many 
\emph{physically} equivalent or only mathematically related. In other words,
the problem is whether the physical information contained in the theory is
``preserved'' or not by the conformal transformations. In other words, one has
the metric $g_{ij} $ and its conformally related one $\bar{g}_{ij}$: the
question is what is  the ``physical metric'', \emph{i.e.}, the metric from
which curvature, geometry, and physical effects have to be calculated and
compared with experiments and observations \cite{mercadante}. 

More precisely, every Killing  and 
homothetic vector  is also  a symmetry of the 
energy-momentum tensor. The latter is also a {\it metric}
of spacetime because a "metric" can be considered   any second order tensor.  
Therefore, up to the homothetic group, metric, energy-momentum tensor, and Ricci tensor have
 the same symmetries. This is not the case for the conformal group and then the requirement 
that the theory is invariant under conformal transformations is an additional assumption not related to 
the gauge invariance.  Furthermore, we are concerned with Lagrangians which means that we assume the 
same kinematics on which these Lagrangians set up a dynamics. In general, kinematics has other symmetries 
with respect to dynamics and these symmetries are not related to those of the metric because no field equation 
relates kinematic quantities to metric. Kinematic symmetries are constrained only by the Ricci identity which gives the constraint and propagation equations.

The issue of
``which is the physical frame'' has been debated for a long time and it
emerges as soon as some authors argue in favor of one frame against the other,
and others support the idea that the two frames are physically equivalent. In
the latter case, authors claim that the issue is a pseudo-problem. The final
result is that there is a good deal of confusion in the literature.

Fierz was the first to pose the problem
\cite{GrumillerKummerVassilevich03}, but the main argument is due to Dicke,
who discussed the conformal transformation for Brans-Dicke theory
\cite{Dicke62}. The point was that physics must be invariant under a rescaling
of units and the conformal transformation is a local rescaling: units do not
change rigidly over the entire spacetime manifold, but by amounts which are
different at different spacetime points. From Dicke's point of view, the two
frames are equivalent provided that mass, length, time, and quantities
derived from them scale with appropriate powers of the conformal factor in the
Einstein frame \cite{Dicke62}.

From this point of view, it is not difficult to see why many authors
consider the issue of which is the physical frame a pseudo-problem. In
principle, it is difficult to object to this argument, but there are some
difficulties.

Even though the above argument is clear in principle, its application
to practical situations gives rise to problems. The assumption that the two
conformal frames are just different \emph{representations} of the same theory,
similar to different gauges of a gauge theory, has to be checked explicitly by
using the field equations of a given system. \textquotedblleft Physical
equivalence\textquotedblright\ is a vague concept because one can consider
many different matter (or test) fields in curved spacetime and different types
of physics, or different physical aspects of a problem. When checking
explicitly the physical equivalence between the two frames, one has to specify
which physical field, or physical process is considered and the equations
describing it. The equivalence has to be shown explicitly, but there is no
proof that holds in any situation (e.g. scalar fields, spinors, cosmology,
black holes derived from the same theory). While physical equivalence can be
proven for various physical aspects, no proof comprehensive of all physical
fields and different physical applications exists.

It is important to stress that Dicke seems  to mix the concept of dimensional units and the concept
of measuring units. For example, spatial distance has the  dimensional unit length $L$
but the measuring unit  length, as we already know from Special
Relativity,  has to be defined in a relativistic inertial frame (RIF). After,
 by some rule,  spatial lengths are compared in different RIFs.  One of such
rules is the Einstein one.  It can be defined by means of light signals which are
simultaneous in the RIF in which the measurement is taking place. This
is the so called {\it rest length} (which, by assumption, coincides with the
corresponding Euclidian length measured by a Newtonian inertial observer).
This type of measurement has nothing to do with the concept of dimension
which, for all RIFs, is the same i.e. the time $T$ (the second, in units,  where $c=1$, see e.g. \cite{mikebook}).
    Dicke is aware of that and then  says {\it " Generally
there may be more than one feasible way of establishing the equality of
units at different spacetime points"} \cite{Dicke62}. It is essential to stress that 
such an approach   is based  on  conventions which may lead to absurd results in real
measurements. Geometrically it is a 1:1 map of two points defining a
spacelike interval from the rest space of one observer at a spacetime point
to a spacelike plane   of another observer at another spacetime point. This
definition it is not a coordinate transformation and there is no point  or
meaning to consider symmetry invariance with this transformation. Dicke  says  {\it  "It is
evident that the equations of motion of matter must be invariant under a
general coordinate dependent transformation of units"}  \cite{Dicke62}. This is a
 misunderstanding that can give rise to confusion. The method of comparing lengths at
different spacetime points is a kinematic assumption while  equations of
motion give dynamics.

Furthermore, Dicke's argument is purely classical. In cosmology, black
hole physics, and quantum fields in curved space, the equivalence of conformal
frames is not clear. At quantum level, this equivalence is not proven due to
the lack of a definitive quantum gravity theory: in fact, when the metric
$g_{ij}$ is quantized, inequivalent quantum theories can be found
\cite{GrumillerKummerVassilevich03,AshtekarCorichi03}. One can consider the
semiclassical regime in which gravity is classical and matter fields are
quantized: again, one would expect the conformal frames to be inequivalent
because the conformal transformations can be seen as Legendre transformations
\cite{MagnanoSokolowski94}, similar to the Legendre transformations of 
classical mechanics of point particles which switches from the canonical
Lagrangian coordinates $q$ to the variables $\left\{  q,p\right\}  $ of the
Hamiltonian formalism. Now, it is well known that Hamiltonians that are
classically equivalent become inequivalent when quantized, producing different
energy spectra and scattering amplitudes \cite{CalogeroDegasperis04,
GlassScanio77}. However, the conformal equivalence between Jordan and Einstein
frame seems to hold to some extent at the semi-classical level
\cite{Flanagan04b}. Again, only a particular kind of physics has been
considered and one cannot make statements about all possible physical
situations.

It is important to point out a very basic argument among particle
physicists that relies on the equivalence theorem of Lagrangian field theory.
It states that the $S$-matrix is invariant under local (nonlinear) field
redefinitions \cite{Dyson48, Blasietal99}. Since the conformal transformation
is, essentially, a field redefinition, it would seem that quantum physics is
invariant under the change of the conformal frame. However, the field theory
in which the equivalence theorem is derived applies to gravity only in the
perturbative regime in which the fields deviate slightly from the Minkowski
space-time. In this regime, tree level quantities can be calculated in any
conformal frame with  same result, but in the non-perturbative regime field
theory and the equivalence theorem do not apply.

Unfortunately, the scaling of units in the Einstein frame often
produces results that either do not make sense or are incorrect  
in the Jordan frame,
 reinforcing the opposite view that the two frames are not equivalent.
While Dicke's explanation is very appealing and several claims supporting the
view that the two frames are inequivalent turned out to be incorrect because
they simply neglected the scaling of units in the Einstein frame, one should
not forget that Dicke's argument is not inclusive of all areas of physics and
it is better to check explicitly that the physics of a certain field does not
depend on the conformal representation and not make sweeping statements.
Certain points have been raised in the literature which either constitute a
problem for Dicke's view, or, at least, indicate that this viewpoint cannot be
applied blindly, including the following ones.

For example, massive particles follow time-like geodesics in the Jordan
frame, while they deviate from geodesic motion in the Einstein frame due to a
force proportional to the gradient of the conformal scalar field
\cite{FaraoniNadeauconfo}. Hence, the Weak Equivalence Principle is satisfied
only in the Jordan frame but not in the Einstein frame due to the coupling of
the scalar field to the ordinary matter. Since the Equivalence Principle is
the foundation of any relativistic theory of gravity, this aspect is important
and there are two ways to consider it. One can ask for the two conformal
frames are equivalent also with respect to the Equivalence Principle. This
means that Equivalence Principle is  formulated in a way that depends on the conformal
frame representation. Then, a representation-independent formulation must be
sought for. However, up to now, no definite result exists in this direction.
On the other hand, we can ask for the violation of the Weak Equivalence
Principle in the Einstein frame by saying that the ``physical equivalence'' of
the two frames must be precisely defined and this concept cannot be used
blindly. In fact the Equivalence Principle holds only in one frame but not in
the other. This fact could be used as an argument against the physical
equivalence of the frames. However, the fact that the Equivalence 
Principle holds in a given frame and not in {\it all frames} means 
that it is not a covariant requirement but a kinematic one. 
In other words, the Equivalence Principle could not be sufficient 
to discriminate among conformally related frames.

In the scalar-tensor theories of gravity, the energy conditions are
easily violated in the Jordan frame, but they are satisfied in the Einstein
frame \cite{MagnanoSokolowski94}. This fact does not eliminate singularities
and then the two frames remain equivalent with respect to the singularities
and not with respect to the energy conditions. This difficulty arises because
part of the matter sector of the theory, in the Einstein frame, comes from the
conformal factor; in other words, the conformal transformation mixes matter
and geometric degrees of freedom, which is the source of many interpretational
problems \cite{Cap01,Cap02}. Thus, even if the theory turns out to be
independent of the conformal representation, its interpretation is not.

There are results in cosmology in which the universe accelerates in
one frame but not in the other. From the pragmatic point of view of an
astronomer attempting to fit observational data (for example, type Ia
supernovae data to a model of the present acceleration of the universe), the
two frames certainly do not appear to be ``physically equivalent''
\cite{nodicap2, CapozzielloPrado}.

To approach correctly the problem of physical equivalence under
conformal transformations, one has to compare physics in different conformal
frames at the level of the Lagrangian, of the field equations, and of their
solutions \cite{allemandi}. This comparison may not always be easy but, in
certain cases, it is extremely useful to discriminate between frames. It has
been adopted, for example, in Ref.~\cite{cno}, to compare cosmological models
in the Einstein and the Jordan frame. Specifically, it has been shown that
solutions of $f(R)$ and scalar-tensor gravity cannot be assumed to be
physically equivalent to those in the Einstein frame when matter fields are
given by generalized Equations of State.

In these and in other situations, one must specify precisely what
``physical equivalence'' means. In certain situations physical equivalence is
demonstrated simply by taking into account the coupling of the scalar field to
matter and the varying units in the Einstein frame, but in other cases the
physical equivalence is not obvious and it does not seem to hold. At the very
least, this equivalence, if it is valid at all, must be defined in precise
terms and discussed in ways that are far from obvious. For this reason, it
would be too simplistic to dismiss the issue of the conformal frame as a
pseudo-problem that has been solved for all physical situations of interest. 

\section{Conformally related metrics and Lagrangians}
Taking in mind the above discussion, we want to seek for 
geometrical structures that are conformally invariant.

 Our aim is to compare cosmological models coming from 
scalar-tensor gravity in order to select conformal quantities  in view of  a possible
physical meaning. This is a very delicate issue that has to be discussed in details.
 On one hand,  an invariant quantity, (i.e. a quantity that remains the same under conformal transformation)  should have a physical meaning.  However,  in cosmology, this 
statement does not necessarily hold since the problem of
equivalence between the two frames is not well posed and it may happen that one of
them has to be taken as the physical one in a particular case and then such
invariant quantity would not have a physical meaning. This situation often happen if cosmological solutions fit data  and then related quantities are assumed as "physical" \cite{CapozzielloPrado}.  On the other hand, if a scalar
field describes an actual particle in a given frame, then its properties (e.g, its
mass and couplings to other fields)  would change  in the
conformally-related metric. This fact does not mean that its properties have no physical meaning.  In other words,  the identification of conformally invariant  physical quantities is a very difficult task if it is not based on first principles.

 With these considerations in mind,
let us start with defining some geometrical  structures that will be useful in the discussion. 

A vector field $X^{a}$ is a Conformal
Killing Vector \cite{Yano} (hereafter CKV) of the metric $g_{ij}$ if there
exists a function $\psi\left(  x^{k}\right)  $ so that:
\begin{equation}
\mathcal{L}_{X}g_{ij}=2\psi\left(  x^{k}\right)  g_{ij}\label{CLN.01}%
\end{equation}
where $\mathcal{L}_{X}$ is the Lie derivative. In case that $\psi_{;a}=0$ i.e.
$\psi=$constant, the vector  $X$ is called homothetic (hereafter HV) while if $\psi=0$
then the vector $X$ is a Killing vector (hereafter KV). In this context,
two metrics $g_{ij},\bar{g}_{ij}$ are said to be conformal or conformally
related if there exists a function $N^{2}\left(  x^{k}\right)  $ so that
$\bar{g}_{ij}=N^{2}\left(  x^{k}\right)  g_{ij}$.
From the mathematical point of view the CKVs form the so called
\textit{conformal algebra} of the metric. The conformal algebra contains two
closed sub-algebras the \textit{Homothetic algebra} and the \textit{Killing
algebra}. Interestingly the above algebras are related as follows:
\begin{equation}
KV_{s}\subseteq HV_{s}\subseteq CKV_{s}\;.\label{CLN.02}%
\end{equation}
The dimension of the conformal algebra of an $n-$ dimensional metric $(n>2)$
of constant curvature equals $\frac{(n+1)(n+2)}{2}$, the dimension of the
Killing algebra $\frac{n(n+1)}{2}$ and that of the homothetic algebra
$\frac{n(n+1)}{2}+1$. Note that two conformally related metrics have the same
conformal algebra \cite{TsampC}, however not the same subalgebras. Indeed if
$X$ is a CKV for the metric $g_{ij}$ i.e. $\mathcal{L}_{X}g_{ij}=2\psi\left(
x^{k}\right)  g_{ij}$ then for the metric $\bar{g}_{ij}$ the vector $X$ is
again a CKV with conformal factor $\bar{\psi}\left(  x^{k}\right)  ,$ that
is:
\begin{equation}
\mathcal{L}_{X}\bar{g}_{ij}=2\bar{\psi}\left(  x^{k}\right)  \bar{g}%
_{ij}\label{CLN.03}%
\end{equation}
where the conformal factors $\psi\left(  x^{k}\right)  $, $\bar{\psi}\left(
x^{k}\right)  $ are related as follows:
\begin{equation}
\bar{\psi}\left(  x^{k}\right)  =\psi\left(  x^{k}\right)  +\mathcal{L}%
_{X}\left(  \ln N\right)  .\label{CLN.04}%
\end{equation}

The Ricci scalars of the conformally related metrics $g_{ij},\bar{g}_{ij}$ are
related as follows \cite{HawkingB}:%
\begin{equation}
\bar{R}=N^{-2}R-2(n-1)N^{-3}\Delta_{2}N-(n-1)(n-4)\Delta_{1}N\label{CLN.04.1}%
\end{equation}
where (note that $\Delta_{2}N$ contains the covariant derivative whereas
$\Delta_{1}N$ the partial derivative):
\begin{align}
\Delta_{1}N &  =g_{ij}N^{,i}N^{,j}\label{CLN.04.2}\\
\Delta_{2}N &  =g_{ij}N^{;ij}.\label{CLN.04.3}%
\end{align}
From the above discussion it becomes clear that all two dimensional spaces are
Einstein Spaces $\left(  \text{i.e. }R_{ab}=\frac{R}{2}g_{ab}\right)  $ and
conformally flat. The metric of a two dimensional space can be written in the
generic form:
\begin{equation}
ds^{2}=N^{2}\left(  x,y\right)  \left(  \varepsilon dx^{2}+dy^{2}\right)
~~,~\varepsilon=\pm1\;.\label{CLN.10}%
\end{equation}

\subsection{Conformal Lagrangians}

Due to the fact that almost every dynamical system is described by a
corresponding Lagrangian, below we study generically, as much as possible, the
problem of the conformal Lagrangians and then we  apply the current ideas
to the scalar field cosmology. To begin with,  consider  the Lagrangian of a
particle moving under the action of a potential $V(x^{k})$ in a Riemannian
space with metric $g_{ij}$
\begin{equation}
L=\frac{1}{2}g_{ij}\dot{x}^{i}\dot{x}^{j}-V\left(  x^{k}\right)  ~,~\dot
{x}^{i}=\frac{dx^{i}}{dt}\label{CLN.05}%
\end{equation}
where $t$ is a path parameter. The equations of motion follow from the action
\begin{equation}
S=\int dxdtL\left(  x^{k},\dot{x}^{k}\right)  =\int dxdt\left[  \frac{1}%
{2}g_{ij}\dot{x}^{i}\dot{x}^{j}-V\left(  x^{k}\right)  \right]
.\label{CLN.06}%
\end{equation}
Changing  the variables in Eq.(\ref{CLN.06}) from $t$ to $\tau$ via the
relation:
\begin{equation}
d\tau=N^{2}\left(  x^{i}\right)  dt \label{tran1A}%
\end{equation}
the action is given by
\begin{equation}
S=\int dx\frac{d\tau}{N^{2}\left(  x^{k}\right)  }\left[  \frac{1}{2}%
g_{ij}N^{4}\left(  x^{k}\right)  x^{\prime i}x^{\prime j}-V\left(
x^{k}\right)  \right]  ~~\label{CLN.07}%
\end{equation}
where $x^{\prime i}=\frac{dx^{i}}{d\tau}$. Obviously, the Lagrangian in the
new coordinate system $(\tau,x^{i})$ becomes:%
\begin{equation}
\bar{L}\left(  x^{k},x^{\prime k}\right)  =\frac{1}{2}N^{2}\left(
x^{k}\right)  g_{ij}x^{\prime i}x^{\prime j}-\frac{V\left(  x^{k}\right)
}{N^{2}\left(  x^{k}\right)  }.\label{CLN.08}%
\end{equation}

Now if we consider a conformal transformation of the metric $\bar{g}%
_{ij}=N^{2}\left(  x^{k}\right)  g_{ij}$ and a new potential function $\bar
{V}\left(  x^{k}\right)  =\frac{V\left(  x^{k}\right)  }{N^{2}\left(
x^{k}\right)  }$ then the new Lagrangian $\bar{L}\left(  x^{k},x^{\prime
k}\right)  $ takes the following form:
\begin{equation}
\bar{L}\left(  x^{k},x^{\prime k}\right)  =\frac{1}{2}\bar{g}_{ij}x^{\prime
i}x^{\prime j}-\bar{V}\left(  x^{k}\right)  \label{CLN.09}%
\end{equation}
implying that Eq.(\ref{CLN.09}) is of the same form as the Lagrangian $L$ in
Eq.(\ref{CLN.05}). From now on \textit{the Lagrangian $L\left(  x^{k},\dot
{x}^{k}\right)  ~$ of Eq.(\ref{CLN.05}) and the Lagrangian $\bar{L}\left(
x^{k},x^{\prime k}\right)  $ of Eq.(\ref{CLN.09}) will be called conformal}.
In this framework, the action remains the same i.e. it is invariant under the
change of parameter, the equations of motion in the new variables $(\tau
,x^{i})$ will be the same with the equations of motion for the Lagrangian $L$
in the original coordinates $(t,x^{k})$ .

It has been shown \cite{TsamGE} that the Noether symmetries of a Lagrangian of
the form (\ref{CLN.05}) follow the homothetic algebra of the metric $g_{ij}.$
The same applies to the Lagrangian $\bar{L}\left(  x^{k},x^{\prime k}\right)
$ and the metric $\bar{g}_{ij}.$ As we have remarked the conformal algebra of
the metrics $g_{ij},\bar{g}_{ij}$ (as a set) is the same however their closed
subgroups of HVs and KVs are in general different\footnote{The Noether
symmetries of the conformal Lagrangians (\ref{CLN.05})-(\ref{CLN.09}) are
elements of the common conformal algebra of the metrics $g_{ij},\bar{g}_{ij}$.
A clear definition of the Noether symmetries can be founds in
\cite{CapRev, Cap01,Cap02,TsamGE} (for applications to cosmology see \cite{BB,felice,nesseris} and
references therein).}. Now, we formulate and prove the following proposition:

\textbf{\emph{Lemma:}} \textit{The Euler-Lagrange equations for two conformal
Lagrangians transform covariantly under the conformal transformation relating
the Lagrangians iff the Hamiltonian vanishes.}

\textbf{\emph{Proof:}} Consider the Lagrangian $L=\frac{1}{2}g_{ij}\dot{x}%
^{i}\dot{x}^{j}-V\left(  x^{k}\right)  $ whose Euler-Lagrange equations are:
\begin{equation}
\ddot{x}^{i}+\Gamma_{jk}^{i}\dot{x}^{i}\dot{x}^{j}+V^{,i}=0 \label{CLN1.01}%
\end{equation}
where $\Gamma_{jk}^{i}$ are the Christofell symbols. The corresponding
Hamiltonian is given by
\begin{equation}
E=\frac{1}{2}g_{ij}\dot{x}^{i}\dot{x}^{j}+V\left(  x^{k}\right)  \;.
\label{CLN1.02}%
\end{equation}
For the conformally related Lagrangian $\bar{L}\left(  x^{k},x^{\prime
k}\right)  =\left(  \frac{1}{2}N^{2}\left(  x^{k}\right)  g_{ij}x^{\prime
i}x^{\prime j}-\frac{V\left(  x^{k}\right)  }{N^{2}\left(  x^{k}\right)
}\right)  $ where $N_{,j}\neq0$ the resulting Euler Lagrange equations are%
\begin{equation}
x^{\prime\prime i}+\hat{\Gamma}_{jk}^{i}x^{\prime j}x^{\prime k}+\frac
{1}{N^{4}}V^{,i}-\frac{2V}{N^{5}}N^{,i}=0 \label{CLN1.03}%
\end{equation}
where
\begin{equation}
\hat{\Gamma}_{jk}^{i}=\Gamma_{jk}^{i}+(\ln N)_{,k}\delta_{j}^{i}+(\ln
N)_{,j}\delta_{k}^{i}-(\ln N)^{,i}g_{jk} \label{CLN1.04}%
\end{equation}
and the corresponding Hamiltonian is%
\begin{equation}
\bar{E}=\frac{1}{2}N^{2}\left(  x^{k}\right)  g_{ij}\dot{x}^{i}\dot{x}^{j}%
+\frac{V\left(  x^{k}\right)  }{N^{2}\left(  x^{k}\right)  }. \label{CLN1.05}%
\end{equation}
In order to show that the two equations of motion are conformally related we
start from Eq.(\ref{CLN1.03}) and apply the conformal transformation
\begin{align*}
x^{\prime i}  &  =\frac{dx^{i}}{d\tau}=\frac{dx^{i}}{dt}\frac{dt}{d\tau}%
=\dot{x}^{i}\frac{1}{N^{2}}\\
x^{^{\prime\prime}i}  &  =\ddot{x}^{i}\frac{1}{N^{4}}-2\dot{x}^{i}\dot{x}%
^{j}\left(  \ln N\right)  _{,j}\frac{1}{N^{4}}.
\end{align*}
Replacing in Eq.(\ref{CLN1.03}) we find:%
\[
\ddot{x}^{i}\frac{1}{N^{4}}-2\dot{x}^{i}\dot{x}^{j}\left(  \ln N\right)
_{,j}\frac{1}{N^{4}}+\frac{1}{N^{4}}\hat{\Gamma}_{jk}^{i}\dot{x}^{j}\dot
{x}^{k}+\frac{1}{N^{4}}V^{,i}-\frac{2V}{N^{5}}N^{,i}=0\,.
\]
Replacing $\hat{\Gamma}_{jk}^{i}$ from Eq.(\ref{CLN1.04}) we have
\begin{align*}
\ddot{x}^{i}-2\dot{x}^{i}\dot{x}^{j}\left(  \ln N\right)  _{,j}+\Gamma
_{jk}^{i}\dot{x}^{j}\dot{x}^{k}+2(\ln N)_{,j}\dot{x}^{j}\dot{x}^{i}\\
-(\ln N)^{,i}g_{jk}\dot{x}^{j}\dot{x}^{k}+V^{,i}-2V(\ln N)^{,i}=0
\end{align*}
from which follows%
\[
\ddot{x}^{i}+\Gamma_{jk}^{i}\dot{x}^{j}\dot{x}^{k}+V^{,i}-(\ln N)^{,i}\left(
g_{jk}\dot{x}^{j}\dot{x}^{k}+2V\right)  =0.
\]
Obviously, the above Euler-Lagrange equations coincide with Eqs.(\ref{CLN1.01}%
) if and only if $\left(  g_{jk}\dot{x}^{j}\dot{x}^{k}+2V\right)  =0$, which
implies that the Hamiltonian of Eq.(\ref{CLN1.02}) vanishes. The steps are
reversible hence the inverse is also true.

The physical  meaning of such a result is that systems with vanishing energy are   conformally  invariant at the level of equations of motion.

\section{Conformally equivalent Lagrangians in the scalar field Cosmology}

Let us  discuss now the conformal equivalence of Lagrangians for scalar
fields in a general Riemannian space of four dimensions (see also
\cite{allemandi}). The field equations in the scalar-tensor  cosmology can be
derived from two different variational principles. In the first case we
consider a scalar field $\phi$ which is minimally coupled to gravity and the
equations of motion follow from the action
\begin{equation}
S_{M}=\int d\tau dx^{3}\sqrt{-g}\left[  R+\frac{\varepsilon}{2}g_{ij}\phi
^{;i}\phi^{;j}-V\left(  \phi\right)  \right]  \label{CLN.11}%
\end{equation}
where
$\varepsilon=\pm 1$ 
defines quintessence or phantom field cosmology respectively.

In the second case we assume a scalar field $\psi$ (different from the
minimally coupled scalar field $\phi$) which interacts with the gravitational
field (non minimal coupling) and the corresponding action is given by
\begin{equation}
S_{NM}=\int d\tau dx^{3}\sqrt{-\bar{g}}\left[  F\left(  \psi\right)  \bar
{R}+\frac{\varepsilon}{2}\bar{g}^{ij}\psi_{;i}\psi_{;j}-\bar{V}\left(
\psi\right)  \right]  \label{CL.12.0}%
\end{equation}
where $F(\psi)$ is the coupling function between the gravitational and the
scalar field $\psi$ respectively. Below we pose the following proposition.

\textbf{\emph{Theorem:}} \textit{The field equations for a non minimally
coupled scalar field $\psi$ with Lagrangian $\bar{L}\left(  \tau,x^{k},\dot
{x}^{k}\right)  $ and coupling function $F(\psi)$ in the gravitational field
$\bar{g}_{ij}$ are the same with the field equations of the minimally coupled
scalar field $\Psi$ for a conformal Lagrangian $L\left(  \tau,x^{k},\dot
{x}^{k}\right)  $ in the conformal metric $g_{ij}$ $=N^{-2}\bar{g}_{ij}$ where
the conformal function is $N=\frac{1}{\sqrt{-2F\left(  \psi\right)  }}$ with
$F\left(  \psi\right)  <0.$ The inverse is also true, that is, to a minimally
coupled scalar field it can be associated a unique non minimally coupled
scalar field in a conformal metric and with a different potential function.}

\textbf{\emph{Proof:}} We first start with the action of Eq.(\ref{CL.12.0}).
Let $g_{ij}$ be the conformally related metric (this is not a coordinate
transformation):
\[
g_{ij}=N^{-2}\bar{g}_{ij}.
\]
Then the action provided by Eq.(\ref{CL.12.0}) becomes \footnote{For a $4\times 4$
matrix namely, $A=(a_{ij})$ we have
\[
\det A=\varepsilon^{ijkl}a_{ij}a_{kl}%
\]
hence
\[
\bar{g}=\varepsilon^{ijkl}\bar{g}_{ij}\bar{g}_{kl}=N^{4}g.\label{CLN.12.1}%
\]
}:%
\[
S_{NM}=\int d\tau dx^{3}N^{4}\sqrt{-g}\left[  F\left(  \psi\right)  \bar
{R}+\frac{\varepsilon}{2}N^{-2}g^{ij}\psi_{;i}\psi_{;j}-\bar{V}\left(
\psi\right)  \right]  .
\]
Inserting the Ricci scalar $\bar{R}$ (using $n=4)$ from Eq.(\ref{CLN.04.1})
into the latter equation we find: \begin{widetext}
\begin{eqnarray}
\label{ICLN}
S_{NM}
&=&\int d\tau dx^{3}N^{4}\sqrt{-g}\left[ F\left( \psi \right)
N^{-2}R-6F\left( \psi \right) N^{-3}\Delta _{2}N
+\frac{\varepsilon }{2}N^{-2}\Delta _{1}\psi -\bar{V}\left( \psi \right) \right]\,.
\end{eqnarray}%
\end{widetext}
Now we can define the conformal function $N$ in terms of the coupling function
$F(\psi)$ [where $F\left(  \psi\right)  <0)$]:
\begin{equation}
\label{CLN122}N=\frac{1}{\sqrt{-2F\left(  \psi\right)  }}~.
\end{equation}
with
\begin{equation}
\label{CLN123}N_{;i}=\frac{F_{\psi}\psi_{;i}}{\left(  -2F\right)  ^{\frac
{3}{2}}}\;.
\end{equation}

Using Eqs.(\ref{CLN122}) and (\ref{CLN123}) the first term of the integral in
Eq.(\ref{ICLN}) becomes:
\[
\int d\tau dx^{3}\sqrt{-g}F\left(  \psi\right)  N^{2}R=\int d\tau dx^{3}%
\sqrt{-g}\left(  -\frac{R}{2}\right)  .
\]
On the other hand the second term in Eq.(\ref{ICLN}) gives, after integration
by parts: \begin{widetext}
\begin{eqnarray}
\label{ICLN1}
\int d\tau dx^{3}\sqrt{-g}\left[ -6F\left( \psi \right) N\Delta _{2}N\right]
&=&\int d\tau dx^{3}\sqrt{-g}\left( -6\frac{F}{\sqrt{-2F}} N_{;ij}g^{ij}\right) \nonumber\\
&=&\int d\tau dx^{3}\sqrt{-g}\left[ -6\frac{F}{\sqrt{-2F}}\frac{1}{\sqrt{-g}}%
\left( \sqrt{-g}g^{ij}N_{,k}\right) _{,j}\right] \nonumber\\
&=&\int d\tau dx^{3}\left[ -6\frac{F}{\sqrt{-2F}}\left( \sqrt{-g}
g^{ij}N_{,k}\right) _{,j}\right] \nonumber\\
&=&\int d\tau dx^{3}\sqrt{-g}\left( 3\frac{F_{\psi }}{\sqrt{-2F}}\psi
_{;j}N_{;i}g^{ij}\right) \nonumber \\
&=&
\int d\tau dx^{3}\sqrt{-g}\left[ 3\frac{F_{\psi }^{2}}{\left( -2F\right)
^{2}}\psi _{;i}\psi _{;j}g^{ij}\right] \;.
\end{eqnarray}
\end{widetext}


The third term provides:
\[
\frac{\varepsilon}{2}N^{2}\Delta_{1}\psi=\frac{\varepsilon}{4F}\psi_{;i}%
\psi_{;j}g^{ij}\,.%
\]
Finally, collecting all terms  and inserting them into Eq.(\ref{ICLN}),
the action is written as \begin{widetext}
\begin{eqnarray}
S_{NM} &=&
\int d\tau dx^{3}\sqrt{-g}\left[ -\frac{R}{2}+3\frac{F_{\psi }^{2}}{4F^{2}
}\psi _{;i}\psi _{;j}g^{ij}-\frac{\varepsilon }{4F}\psi _{;i}\psi
_{;j}g^{ij}-\frac{\bar{V}\left( \psi \right) }{4F^{2}}\right] \nonumber\\
&=&
\int d\tau dx^{3}\sqrt{-g}\left[ -\frac{R}{2}+\frac{\varepsilon }{2}%
\left( \frac{3\varepsilon F_{\psi }^{2}-F}{2F^{2}}\right) \psi _{;i}\psi
_{;j}g^{ij}-\frac{\bar{V}\left( \psi \right) }{4F^{2}}\right] \;.
\label{CLN.12.5}
\end{eqnarray}
\end{widetext}Interestingly, introducing the scalar field $\Psi$ with the
requirement:
\begin{equation}
d\Psi=\sqrt{\left(  \frac{3\varepsilon F_{\psi}^{2}-F}{2F^{2}}\right)  }%
d\psi\label{CLN.12.6}%
\end{equation}
the action of Eq.(\ref{CLN.12.5}) can be written as follows
\begin{equation}
S_{NM}=\int d\tau dx^{3}\sqrt{-g}\left[  -\frac{R}{2}+\frac{\varepsilon}%
{2}\Psi_{;i}\Psi_{;j}g^{ij}-\frac{\bar{V}\left(  \Psi\right)  }{4F\left(
\Psi\right)  ^{2}}\right]  \;.\label{CLN.12.7}%
\end{equation}
We conclude that the scalar field $\Psi$ is minimally coupled (modulus a
constant) to the gravitational field. In other words, we find that to every
non-minimally coupled scalar field, we may associate a unique minimally coupled
scalar field in a conformally related space with an appropriate potential. All
considerations are reversible, hence the result is reversible. Finally, we
would like to remark that the above theorem can be extended to general
Riemannian spaces of $n-$dimensions (see appendix A).

\section{Conformal Lagrangians in FRLW cosmology}

In this section we consider a spatially flat $\left(  K=0\right)  $ FRLW
spacetime\footnote{Similar results can be achieved  for $K\neq0$ \cite{allemandi}.}
whose metric is%
\begin{equation}
ds^{2}=-dt^{2}+a^{2}\left(  t\right)  \delta_{ij}dx^{i}dx^{j}\label{CLN.13}%
\end{equation}
where $\delta_{ij}$ is the 3-space metric in Cartesian coordinates. The
Lagrangian of a scalar field $\phi$ minimally coupled to gravity in this
coordinate system $(a,\phi)$ is
\begin{equation}
L_{M}=-3a\dot{a}^{2}+\frac{\varepsilon}{2}a^{2}\dot{\phi}^{2}-a^{3}V\left(
\phi\right)  .\label{CLN.14}%
\end{equation}
On the other hand, the Lagrangian of the non minimally coupled scalar field
$\psi$ in the coordinate system $(a,\psi)$ is given by
\begin{equation}
L_{NM}=6F\left(  \psi\right)  a\dot{a}^{2}+6F_{\psi}\left(  \psi\right)
a^{2}\dot{a}\dot{\psi}+\frac{\varepsilon}{2}a^{3}\dot{\psi}^{2}-a^{3}V\left(
\psi\right)  \label{CLN.15}%
\end{equation}
where $F(\psi)<0$ is the coupling function. The Hamiltonian of the above
Lagrangian is
\begin{equation}
E=6F\left(  \psi\right)  a\dot{a}^{2}+6F_{\psi}\left(  \psi\right)  a^{2}%
\dot{a}\dot{\psi}+\frac{\varepsilon}{2}a^{3}\dot{\psi}^{2}+a^{3}V\left(
\psi\right)  .\label{CLN.15e}%
\end{equation}

We construct a conformal Lagrangian which corresponds to a minimally coupled
scalar field. To do that we introduce the following transformation (see
\cite{Cap01,Cap02}):
\begin{equation}
A\left(  t\right)  =\sqrt{-2F(t)}a(t)\;.\label{CLN.15a}%
\end{equation}
Then the Lagrangian (\ref{CLN.15}) takes the form:%
\begin{align}
L_{NM} &  =\frac{1}{\sqrt{-2F}}\left[  -3A\dot{A}^{2}+\frac{\varepsilon}%
{2}\left(  \frac{3\varepsilon F_{\psi}^{2}-F}{2F^{2}}\right)  A^{3}~\dot{\psi
}^{2}\right]  \nonumber\\
&  -\frac{A^{3}}{\left(  -2F\right)  ^{\frac{3}{2}}}V\left(  \psi\right)
\;.\label{CLN.16}%
\end{align}
It is interesting to mention here that the cross term $\dot{a}\dot{\psi}$
disappears from Eq.(\ref{CLN.16}). Utilizing simultaneously Eq.(\ref{CLN.12.6}%
) and the conformal transformation
\begin{equation}
d\tau=\sqrt{-2F\left(  \psi\right)  }dt\label{CLN.17a}%
\end{equation}
we find, after some algebra, that
Eq.(\ref{CLN.16}) can be  written as
\begin{equation}
L_{M}\left(  A,A^{\prime},\Psi,\Psi^{\prime}\right)  =-3AA^{\prime2}%
+\frac{\varepsilon}{2}A^{3}\Psi^{\prime2}-A^{3}\bar{V}(\Psi)\label{LLC}%
\end{equation}
where
\begin{equation}
\bar{V}(\Psi)=\frac{A^{3}}{\left(  -2F\right)  ^{\frac{3}{2}}}V\left(
\Psi\right)  \;.\label{CLN.17}%
\end{equation}
Notice that the prime denotes derivative with respect to the conformal time
$\tau$.

Evidently, the functional form of the Lagrangian (\ref{LLC}) has the general
form of Eq.(\ref{CLN.14}) proving our assessment. Furthermore, considering in
the new coordinates $(\tau,x^{i})$ the metric%
\begin{equation}
d\bar{s}^{2}=-d\tau^{2}+A^{2}\left(  \tau\right)  \delta_{ij}dx^{i}%
dx^{j}\label{CLN.19}%
\end{equation}
we find that the term $3AA^{\prime2}$ equals the Ricci scalar $\bar{R}$ of the
conformally flat metric $d\bar{s}^{2}.$ In other words, the Lagrangian
(\ref{LLC}) can be seen as the Lagrangian of a scalar field $\Psi$ of
potential $\bar{V}\left(  \Psi\right)  $ which is minimally coupled to the
gravitational field $\bar{g}_{ij}$ in the space with metric $d\bar{s}^{2}$.
Replacing the quantity $A\left(  \tau\right)  $ and the coordinate $\tau$ from
Eq.(\ref{CLN.15a}) and Eq.(\ref{CLN.17a}) respectively, we obtain:
\begin{equation}
d\bar{s}^{2}=\sqrt{-2F}\left[  -dt^{2}+a^{2}\left(  t\right)  \delta
_{ij}dx^{i}dx^{j}\right]  =\sqrt{-2F}ds^{2}\label{CLN.20}%
\end{equation}
that is, the metric $d\bar{s}^{2}$ is conformally related to the metric
$ds^{2}$ with conformal function $\sqrt{-2F}.$ This means that the
non-minimally coupled scalar field in the gravitational field $ds^{2}$ is
equivalent to a minimally coupled scalar field - with appropriate potential
defined in terms of the coupling function - in the gravitational field
$d\bar{s}^{2}.$ For the benefit of the reader, we would like to stress that the
above geometrical/dynamical result is reversible in the sense that a minimally
coupled scalar field $\phi$ in a metric $ds^{2}$ can be seen as a non-minimally coupled scalar field $\psi$ in the flat FRLW space in which the
Eq.(\ref{CLN.15}) is equivalent to the minimally coupled scalar field
$\Psi=\Psi(\psi)$ in the conformally related metric $d\bar{s}^{2},$ where the
conformal function is defined in terms of the coupling function. Equivalently
the Lagrangians $L_{M}$ and $L_{NM}$ are conformally related. 
Finally, we want to stress that the result of the previous lemma is automatically recovered since 
the Hamiltonian (\ref{CLN.15e}) is equal to zero being the $\{0,0\}$ Einstein equation of the system (see also \cite{CapRev}).

\section{Discussion and Conclusions}

In this article we have investigated  conformally related metrics and Lagrangians
in the context of scalar-tensor cosmology.  The aim is to select which is the frame where conformally related solutions have an immediate  physical meaning.
As  discussed in section II, no final statement is available for the problem if solutions have to be interpreted either in the Jordan frame or in the Einstein frame since the physical equivalence  can be questioned according to several issues (quantum vs classical measurements, energy conditions,  choice of physical units, etc.).
Due to this situation, it is is too simplistic to consider the problem of conformal frames just a pseudo-problem since we are facing only a mathematical equivalence.

Clearly, it has to be addressed at three levels: $i)$ Lagrangians (or in general, effective actions); $ii)$ field equations; $iii)$ solutions. Actually, the last issue means also the choice of a set of  observables where the interpretation of solutions is evident. To this goal, seeking for dynamical quantities invariant under conformal transformations is a fundamental issue. However, such quantities have to be related to geometry and possibly to be conserved like Noether symmetries.

With this target in mind, we have firstly proved a lemma which
shows that the field equations of two conformally related Lagrangians are also
conformally related if the corresponding Hamiltonian vanishes. This fact is extremely relevant being the Hamiltonian the energy constraint of a given mechanical system and, in particular, it constitutes a non-holonomic constraint for dynamical systems describing cosmological models. It is  the $\{0,0\}$ Einstein equation of the system.

Secondly, we have found that to every non-minimally coupled scalar field, we can associate a
unique minimally coupled scalar field in a conformally related space with an
appropriate potential. The existence of such a connection can be used in order
to study the dynamical properties of the various cosmological models, since
the field equations of a non-minimally coupled scalar field are the same, at
the conformal level, of  the field equations of the minimally coupled scalar
field. The above propositions can be extended to  general
Riemannian spaces in  $n$-dimensions.

It is worth stressing that the above results are in agreement with the so called {\it Bicknell's Theorem} which  states that  a general non-linear $f(R)$ Lagrangian is equivalent to a minimally coupled scalar field with a general potential in the Einstein frame. In Ref. \cite{Bick74}, this result is achieved in the case of $R^2$-gravity. In \cite{schmidt}, the result is generalized to any analytic $f(R)$-gravity.
We'd like to point out
that in a recent paper \cite{BB}, based on the Noether symmetry approach,  
we have studied the issue of physical solutions in  $f(R)$ gravity models and scalar field dark energy models.
Starting from these results, it is possible to identify the 
Noether symmetries, the physical solutions and the corresponding
conformal properties of the scalar tensor theories (including $f(T)$ gravity).
Such an analysis is in progress.

In general, the Noether symmetries play an important role in
physics because they can be used to simplify a given system of differential
equations as well as to determine the integrability of the system. The latter
will provide the necessary platform in order to solve the equations of motion
analytically and thus to obtain the evolution of the physical quantities. In  cosmology, such a method is extremely relevant in order to compare cosmographic parameters, such as  scale factor,   Hubble expansion
rate, deceleration parameter, density parameters with observations \cite{CapRev,BB, felice,nesseris}.

\appendix{\section{Generalization of the Theorem to $n$-dimensions}}

In this appendix we generalize the theorem of section III to a Riemannian
space of dimension $n$. Briefly, we consider the non minimally coupled scalar field $\psi $ whose field equations are obtained from the action:
\begin{widetext}
\begin{eqnarray}
S_{NM} &=&\int dx^{n}N^{n}\sqrt{-g}\left[ F\left( \psi \right) \bar{R}+\frac{%
\varepsilon }{2}N^{-2}g^{ij}\psi _{,i}\psi _{,j}-\bar{V}\left( \psi \right) %
\right] \nonumber \\
&=&\int dx^{n}\sqrt{-g}\left[
\begin{array}{c}
F\left( \psi \right) N^{n-2}R-2(n-1)F\left( \psi \right) N^{n-3}\Delta _{2}N+
\\
-F\left( \psi \right) N^{n}(n-1)(n-4)\Delta _{1}N+\frac{\varepsilon }{2}%
N^{n-2}g^{ij}\psi _{,i}\psi _{,j}-N^{n}\bar{V}\left( \psi \right)%
\end{array}%
\right]
\label{GENA}
\end{eqnarray}
\end{widetext}
where in order to derive the last equality we have used Eq.(\ref{CLN.04.1}).
Note that we can define the function $N(x^{i})$ in terms of the coupling
function $F(\psi)$ by the requirement:
\[
N^{n-2}=-\frac{1}{2F}\;\;\;\;F=-\frac{N^{2-n}}{2}%
\]
which also implies that
\begin{equation}
N=\frac{1}{\left(  -2F\right)  ^{\frac{1}{n-2}}}\rightarrow N^{-1}=\left(
-2F\right)  ^{\frac{1}{n-2}} \label{N11}%
\end{equation}
\begin{equation}
N_{;i}N_{;j}^{-1}=-\frac{F_{\psi}^{2}}{\left(  n-2\right)  ^{2}F^{2}}\psi
_{;i}\psi_{;j}\;. \label{N12}%
\end{equation}
We start now to treat the terms of the action in Eq.(\ref{GENA}).

In particular, the first term gives:
\begin{equation}
\int dx^{n}\sqrt{-g}\left(  F\left(  \psi\right)  N^{n-2}R\right)  =\int
dx^{n}\sqrt{-g}\left(  -\frac{R}{2}\right)  \;.\label{ICLN01}%
\end{equation}
If we utilize Eqs.(\ref{N11}) and (\ref{N12}) then 
the second (integrating by parts) and the third terms of the general action are 

\begin{widetext}
\begin{eqnarray}
\label{ICLN11}
\int dx^{n}\sqrt{-g}\left[ -2(n-1)F\left( \psi \right)
N^{n-3}N_{;ij}g^{ij}\right] &=&\int dx^{n}\sqrt{-g}\left[
(n-1)N^{2-n}N^{n-3}N_{;ij}g^{ij}\right] \nonumber \\
&=&\int dx^{n}\sqrt{-g}\left[ (n-1)N^{-1}\frac{1}{\sqrt{-g}}\left( \sqrt{-g}%
g^{ij}N_{,k}\right) _{,j}\right] \nonumber \\
&=&\int dx^{n}\sqrt{-g}\left[ (n-1)N^{-1}N_{;ij}g^{ij}\right] \nonumber \\
&=&\int dx^{n}\sqrt{-g}\left[ -(n-1)\left( N^{-1}\right)
_{;j}N_{;i}g^{ij}\right] \nonumber \\
&=&\int dx^{n}\sqrt{-g}\left[ \frac{(n-1)}{\left(
n-2\right) ^{2}}\frac{F_{\psi }^{2}}{F^{2}}\psi _{;i}\psi _{;j}g^{ij}\right]\;.
\end{eqnarray}
\end{widetext}

\begin{widetext}
\begin{eqnarray}
\label{ICLN22}
\int dx^{n}\sqrt{-g}\left[ F\left( \psi \right) N^{n}(n-1)(n-4)\Delta
_{1}N\right) &=&\int dx^{n}\sqrt{-g}\left( -\frac{N^{2-n}}{2}%
N^{n}(n-1)(n-4)\Delta _{1}N\right] \nonumber \\
&=&\int dx^{n}\sqrt{-g}\left[ -\frac{1}{2}N^{2}(n-1)(n-4)\Delta _{1}N\right] \nonumber \\
&=&\int dx^{n}\sqrt{-g}\left[ -\frac{1}{2}\frac{(n-1)(n-4)}{\left(
n-2\right) ^{2}}\frac{F_{\psi }^{2}}{\left( -2F\right) ^{\frac{4}{2-n}}F^{2}}%
\psi _{;i}\psi _{;j}g^{ij}\right] \;.
\end{eqnarray}
\end{widetext}

To this end the final term gives:%
\begin{widetext}
\begin{equation}
\label{ICLN33}\int dx^{n}\sqrt{-g}\left(  \frac{\varepsilon}{2}N^{n-2}%
g^{ij}\psi_{,i}\psi_{,j}\right)  =\int dx^{n}\sqrt{-g}\left(  -\frac
{\varepsilon}{4}\frac{1}{F}g^{ij}\psi_{;i}\psi_{;j}\right).
\end{equation}
\end{widetext}

Now we change the variable $\psi$ to $\Psi$ as follows
 \begin{widetext}
\begin{equation}
\label{ICLN44}
d\Psi =\left[ \frac{2\varepsilon (n-1)}{\left( n-2\right) ^{2}}\frac{F_{\psi
}^{2}}{F^{2}}-\varepsilon \frac{(n-1)(n-4)}{\left( n-2\right) ^{2}}\frac{%
F_{\psi }^{2}}{\left( -2F\right) ^{\frac{4}{2-n}}F^{2}}-\frac{1}{2F}\right]
^{\frac{1}{2}}d\psi \;.
\end{equation}
\end{widetext}

Collecting the results of the above terms namely Eqs.(\ref{ICLN01}%
),(\ref{ICLN11}),(\ref{ICLN22}),(\ref{ICLN33}) and (\ref{ICLN44}) we find
after some non-trivial algebra that the general action of Eq.(\ref{GENA}) is
written is terms of $\psi$ as follows
\[
S_{NM}=\int dx^{n}\sqrt{-g}\left[  -\frac{R}{2}+\frac{\varepsilon}{2}\Psi
_{;i}\Psi_{;j}g^{ij}-\frac{\bar{V}\left(  \Psi\right)  }{\left(  -2F\right)
^{\frac{n}{n-2}}}\right]  .
\]
We would like to remind the reader that the new scalar field $\Psi$ is
minimally coupled to the gravitational field $g_{ij}$ and that the potential
of $\Psi$ is given by $\frac{\bar{V}\left(  \Psi\right)  }{\left(  -2F\right)
^{\frac{n}{n-2}}}$. Notice that for $n=4$ the above expressions boil down to
those of section III as they should.
The above proof agrees with that provided by Keiser \cite{Keiser} however in
our work we have used a different methodology which is simple and transparent.


\end{document}